\documentclass[twocolumn,aps,pra,showpacs,amsfonts,amsmath,amssymb,superscriptaddress,footinbib, floatfix,longbibliography]{revtex4-1}

\usepackage[T1]{fontenc}
\usepackage{color}
\usepackage{graphicx}
\usepackage[normalem]{ulem}

\usepackage{filemod}

\def\vecr{\mathbf{r}}
\def\vecq{\mathbf{q}}
\def\rmd{\mathrm{d}}
\def\rme{\mathrm{e}}
\def\vecD{\mathbf{D}}
\begin{document}

\title{Multifractality of the kicked rotor at the critical point of the Anderson transition}

\author{Panayotis Akridas-Morel} 
\affiliation{Laboratoire Kastler Brossel, Sorbonne Universit\'e, CNRS, ENS-PSL Research University, Coll\`ege de France, 4 Place Jussieu, 75005 Paris, France}

\author{Nicolas Cherroret} 
\affiliation{Laboratoire Kastler Brossel, Sorbonne Universit\'e, CNRS, ENS-PSL Research University, Coll\`ege de France, 4 Place Jussieu, 75005 Paris, France}

\author{Dominique Delande}
\email[Corresponding author: ]{Dominique.Delande@lkb.upmc.fr}
\affiliation{Laboratoire Kastler Brossel, Sorbonne Universit\'e, CNRS, ENS-PSL Research University, Coll\`ege de France, 4 Place Jussieu, 75005 Paris, France}


\begin{abstract}
We show that quantum wavepackets exhibit a sharp macroscopic peak as they spread in the vicinity of the critical point of the Anderson transition. 
The peak gives a direct access to the mutifractal properties of the wavefunctions and specifically to the multifractal dimension $d_2$. 
Our analysis is based on an experimentally realizable setup, the quantum kicked rotor with quasi-periodic temporal driving, 
an effectively 3-dimensional disordered system recently exploited to explore the physics of the Anderson transition with cold atoms.
 
\end{abstract}

\maketitle

\section{Introduction}
In the vicinity of the critical point of a continuous phase transition, large fluctuations are observed~\cite{Landau:StatisticalPhysics}, responsible for dramatic phenomena like e.g. critical opalescence.
Beyond mean-field descriptions, renormalization group approaches make it possible to describe critical phenomena at (almost) all scales, and to predict  critical exponents~\cite{Wilson:RMP1983}.
Large fluctuations arise as well in quantum phase transitions, where they are usually probed via correlation functions or transport properties. The metal-insulator Anderson transition, taking place in disordered quantum systems, is especially interesting. It separates 
a metallic phase at weak disorder, where transport is diffusive, and an insulating phase at strong disorder, where transport is inhibited due to  interference
in multiple scattering from random defects~\cite{Anderson:LocAnderson:PR58,Ramakrishnan:DisorderElectrons:RMP85}, a phenomenon known as Anderson localization. The dimensionality of the system is a crucial parameter: Anderson localization is the generic scenario in one-dimensional (1D) systems, while the Anderson transition can be observed in 
dimension strictly larger than two. In three-dimensional (3D) systems, the critical point of the Anderson transition occurs for strong disorder, when the product of
the wavenumber $k$ with the mean free path $\ell$ is close to unity, $(k\ell)_c \approx 1$~\cite{Ioffe}.
Although the order parameter for the Anderson transition remains debated~\cite{Evers:AndersonTransitions:RMP2008}, there is nowadays a wide consensus that it is a second order continuous transition, with an algebraic divergence of the localization 
length on the localized side, $\xi\propto 1/((k\ell)_c -k\ell)^\nu$, and an algebraic vanishing of the diffusion coefficient on the diffusive side, $D\propto (k\ell-(k\ell)_c)^s$. 
Numerous evidence for these properties have been found in numerical simulations of the standard
Anderson model, which has been also used to compute the critical exponents $\nu\!=\!s\!\approx\!1.57$ in dimension $d\!=\!3$~\cite{Slevin:CriticalExponent:NJP2014,Slevin:CriticalExponentAndersonTrParallel:JPSP18}. 
This value is universal (depending only on the dimension and symmetry properties) and has been confirmed on other models~\cite{Ghosh:CBS3D:PRL15}. 
Numerical studies of the Anderson model have shown that the distribution of conductance at the critical point is  universal as well, scale invariant~\cite{Markos:UniversalityConductance:EPL1994} and broad, a clear-cut manifestation of large fluctuations at the critical point. 
Large fluctuations also show up in the critical eigenstates, which
are strongly multifractal, displaying regions where $|\psi|^2$ is unexpectedly large and regions where it is unexpectedly small~\cite{Rodriguez:MultifractalFiniteSizeScaling:PRB11}.
Usually, this property is quantitatively described using the generalized inverse participation ratio (GIPR)~\cite{Bell:Atomic,Castellani:multifractal:1986}:
\begin{equation}
 \label{Eq:GIPR}
P_q = \int_{L^d}
\rmd^d\vecr \ |\psi(\vecr)|^{2q},
\end{equation}
where $q$ is a real number and $L$ is the system size. 
The multifractality analysis studies
how the GIPR averaged over eigenstates and/or disorder realizations scales with  $L$. 
If the average -- denoted by $\overline{\phantom{a}}$ in the following -- $\overline{P_q}$ scales like $L^{-\tau_q},$ $\tau_q$ is called the multifractal exponent. By construction $\tau_0\!=\!-d$ and, by normalization of the wavefunction, $\tau_1\!=\!0.$ One can equivalently use the set of multifractal dimensions $d_q=\tau_q/(q-1).$ 
A wavefunction delocalized over a set of dimension $\mathcal{D}$ (which can be an ordinary or a fractal set) will have $d_q=\mathcal{D}$ for all $q$. 
For multifractal states, finally, $d_q$ is a continuous function of $q$ with large positive $q$ values probing the regions of large $|\psi|^2$ and negative $q$ the regions where  $|\psi|^2$ is vanishingly small.

How to experimentally access multifractal dimensions is far from obvious. This in principle requires to measure wavefunctions for various disorder realizations or energies close to the critical point  
\textit{everywhere} in space, a tremendously difficult task.
Alternatively, one can extract only a part of the information on multifractality by selective, less complete, measurements~\cite{Faez:Multifractality:PRL09,Richardella:Multifractality:Science2010,Maschek:Multifractality:NP2012}, for example of the intensity distribution on the exit plane of a disordered slab. 
In this paper, we show that, at long times, the average density of an initially localized wavepacket spreads, but develops near its initial location a sharp peak giving a direct access to the multifractal properties of the wavefunctions and specifically to the multifractal dimension $d_2$. We base our analysis upon an experimentally existing system, the quasi-periodically kicked rotor, which has been shown
to display the Anderson metal-insulator transition~\cite{Chabe:Anderson:PRL08,Lopez:ExperimentalTestOfUniversality:PRL12}, but the mechanism to extract
a multifractal dimension from the expansion of a wavepacket is quite general and could be used in other critical systems. Importantly, although the system is 1D, it can be mapped on a 3D disordered system, so that the measured multifractal dimension $d_2$ is truly the one of 3D critical Anderson-like systems.

The paper is organized as follows. In section~\ref{Sec:KR}, we describe the quasi-periodically kicked rotor and its connection with quantum disordered systems and the metal-insulator Anderson transition. The theoretical approach explaining the origin of the sharp peak in the average density and its connections with multifractality is developed in section~\ref{Sec:Theory}. This knowledge is used in section~\ref{Sec:Measurement}, showing how to extract the multifractal dimension $d_2.$ It is demonstrated on the results of numerical simulations, where the values of the parameters are chosen like in the real experiments. In section~\ref{Sec:Summary}, we briefly discuss the experimental perspectives and summarize our results. 
   
\section{The Quasi-Periodically Kicked Rotor}
\label{Sec:KR}
\subsection{Hamiltonian}
\label{sec:QPKR}
The quasi-periodically kicked rotor (QPKR) is a simple 1D system -- a standard rotor -- exposed periodically to kicks whose amplitude is modulated quasi-periodically in time. With two quasi-periods in addition to the period of the kicks, the Hamiltonian reads:
\begin{eqnarray}
\lefteqn{H=\frac{p^{2}}{2}+K\cos x}\nonumber \\
& &\times \left[1+\varepsilon\cos(\omega_{2}t+\varphi_2)\cos(\omega_{3}t+\varphi_3)\right]\sum_n\delta(t-n)\;,\label{eq:H}
\end{eqnarray}
where the unit of time is the interval between two consecutive kicks.
$K,\varepsilon,\omega_2,\omega_3, \varphi_2, \varphi_3$ (and the Planck's constant $\hbar$ governing the quantum evolution) are dimensionless parameters whose roles are discussed below.
In the limiting case $\varepsilon=0,$ one recovers the usual kicked rotor~\cite{Chirikov:ChaosClassKR:PhysRep79,Izrailev:LocDyn:PREP90}.

As discussed in~\cite{Casati:IncommFreqsQKR:PRL89,Shepelyansky:Kq:PD83,Lemarie:AndersonLong:PRA09}, the dynamics of the QPKR can be mapped on the dynamics of a 3D periodically kicked pseudo-rotor~\footnote{We use the term ``pseudo-rotor'' because of the unusual linear dependence of the Hamiltonian on momenta $p_2$ and $p_3$} with Hamiltonian:
\begin{eqnarray}
\lefteqn{\mathcal{H}=\frac{p_{1}^{2}}{2}+\omega_{2}p_{2}+\omega_{3}p_{3}}\nonumber \\
&  & +K\cos x_{1}\left[1+\varepsilon\cos x_{2}\cos x_{3}\right]\sum_{n}\delta(t-n)\;.
\label{Eq:KR3DquasiperH}
\end{eqnarray}   
More precisely, it is shown in~\cite{Lemarie:AndersonLong:PRA09} that the evolution of 
\begin{equation}
\label{Eq:Psi3}
\Psi({x}_{1},{x}_{2},{x}_{3},t)\equiv\psi({x}_{1},t)\delta({x}_{2}-\varphi_{2}-\omega_2 t)\delta({x}_{3}-\varphi_{3}-\omega_3 t)
\end{equation}
under Hamiltonian $\mathcal{H},$ Eq.~(\ref{Eq:KR3DquasiperH}), is \emph{strictly} equivalent to the evolution of $\psi(x,t)$ under Hamiltonian~$H$, Eq.~(\ref{eq:H}), for any arbitrary initial wavefunction $\psi(x,t\!=\!0)$.

Provided $\omega_2,\omega_3,\pi,\hbar$  are mutually incommensurate real numbers, the 3D periodically kicked pseudo-rotor
can itself be mapped~\cite{Lemarie:AndersonLong:PRA09}
on an anisotropic 3D Anderson model, where $K$ controls the disorder strength and $\varepsilon$ the anisotropy~\cite{Lopez:PhaseDiagramAndersonQKR:NJP13}, a fact
further confirmed by a low-energy effective field theory~\cite{Tian:TheoryAndersonTransition:PRL11}.
An important consequence of these mappings is that the QPKR is a time-dependent 1D system equivalent to a 3D disordered time-independent system, and thus makes it possible to explore the dynamics of the latter, which may display the metal-insulator Anderson transition.
 
\subsection{Dynamical localization}
Depending on the values of the $\hbar,K,\varepsilon,\omega_2,\omega_3, \varphi_2, \varphi_3$ parameters, the 3D disordered system may be localized (at small disorder) or diffusive (at large disorder). The two regimes are separated by a critical point -- usually called mobility edge -- where multifractality is expected to play an important role.
The mapping of the QPKR to the 3D disordered system is discussed in details in~\cite{Lemarie:AndersonLong:PRA09}. An important point is that the statistical properties of the 3D disordered system depend only on the parameters $\hbar,K,\varepsilon.$ Various values of $\omega_2,\omega_3, \varphi_2, \varphi_3$ correspond to various realizations of the disorder, all with the same statistical properties. Hence, one can explore the 3D metal-insulator Anderson transition by varying $\hbar,K,\varepsilon$ with the QPKR.

An important property of the QPKR is  that the localized/delocalized dynamics takes place \emph{in momentum space}, not in position space like the usual Anderson model. Such a localization in momentum space has been dubbed ``dynamical localization''~\cite{Casati:LocDynFirst:LNP79,Fishman:LocDynAnders:PRL82,Moore:LDynFirst:PRL94}. Specifically, 
the system is localized at small $K$ value, i.e. $\overline{\langle p^2(t)\rangle}$ tends to a constant at long $t$, and diffusive at large $K$ value, i.e. $\overline{\langle p^2(t)\rangle} \propto t$ at long $t$ [where $\langle p^2(t)\rangle\equiv\langle\psi(t)|p^2|\psi(t)\rangle$ denotes the quantum mechanical expectation value].
In between, there is a critical point, whose position can be approximately predicted by a mean-field approach, the self-consistent theory of localization (SCTL)~\cite{Lopez:PhaseDiagramAndersonQKR:NJP13,Vollhardt:SelfConsistentTheoryAnderson:92}, where the system behaves sub-diffusively:  $\overline{\langle p^2(t)\rangle} \propto t^{2/3}$ at long $t$. Experimentally, these properties have been confirmed by monitoring the temporal expansion
of a wavepacket initially localized in momentum space around $p=0,$ that is by measuring $|\psi(p,t)|^2$ at increasing time,
with the initial state $|\psi(p,0)|^2\approx \delta(p)$.

In the simplest case of the standard kicked rotor where $\varepsilon=0,$ the 3D disordered system reduces to a 2D array of uncoupled 1D disordered systems, which are localized for all $K$ values. This dynamical localization has been observed experimentally using cold atoms in Ref.~\cite{Moore:AtomOpticsRealizationQKR:PRL95}. The full 3D Anderson transition with $\varepsilon \neq 0$ has been later observed and characterized~\cite{Chabe:Anderson:PRL08,Lemarie:CriticalStateAndersonTransition:PRL10}.

\subsection{Critical point}
At the critical point, the spatial fluctuations of the wavepacket have been numerically and theoretically studied~\cite{Georgeot:Multifractality:PRE2010,Georgeot:Multifractality:PRE2012} from an analysis of the GIPRs, Eq.~(\ref{Eq:GIPR}): they display only very weak multifractal properties. In contrast, we will show in section~\ref{Sec:Theory} that the average density (averaged over disorder realizations) $\overline{|\psi(p,t)|^2}$ itself presents a direct, macroscopic signature of the multifractality of the 3D critical model.

Because of sub-diffusion at the critical point, the width of the wavepacket increases like $t^{1/3},$ but its  global shape is independent of time, a manifestation of scale invariance.
The SCTL makes a definite prediction for this shape~\cite{Lemarie:CriticalStateAndersonTransition:PRL10,Lemarie:These:09}:     
\begin{equation}
\overline{|\psi(p,t)|^2} =\frac{3}{2}\left(3\rho^{3/2}t\right)^{-1/3}\text{Ai}\left[\left(3\rho^{3/2}t\right)^{-1/3}\vert p\vert\right]\;,
\label{Eq:fonccritAi}
\end{equation}
where $\rho=\Gamma(2/3)\Lambda_{c}/3$ is  related to the critical quantity
$\Lambda_{c}=\underset{t\rightarrow\infty}{\lim}\overline{\langle p^{2}(t)\rangle}/t^{2/3},$ a numerical factor depending on the anisotropy $\varepsilon$~\cite{Lopez:PhaseDiagramAndersonQKR:NJP13},
with $\Gamma$ the Gamma function and $\text{Ai}(x)$ the Airy function.
It is convenient to define scaled variables: $\mathcal{P}=pt^{-1/3},\ \mathcal{N}(\mathcal{P},t)=t^{1/3}\ \overline{|\psi(p,t)|^2}$ so that the SCTL prediction reads:
\begin{equation}
\mathcal{N}(\mathcal{P}) = \frac{3^{2/3}}{2\rho^{1/2}}\ \text{Ai}\left(\frac{|\mathcal{P}|}{3^{1/3}\rho^{1/2}}\right),
\label{eq:airyfunc}
\end{equation}
independent of time.
This prediction has been found in excellent agreement with the experimental results on the atomic
 QPKR after few tens of kicks~\cite{Lemarie:CriticalStateAndersonTransition:PRL10}, describing both the kink around $\mathcal{P}=0$ and the tail 
$\propto \exp \left(-\alpha |\mathcal{P}|^{3/2}\right).$

\subsection{Numerical simulations}
In the following, we will use numerical simulations of the QPKR.
The structure of the Hamiltonian of the QPKR, Eq.~(\ref{Eq:KR3DquasiperH}), makes it very easy to numerically propagate any initial state. The free evolution operator between two consecutive kicks is diagonal in the momentum eigenbasis, while the instantaneous kick operator is diagonal in the position eigenbasis.
Because the Hamiltonian is spatially periodic with period $2\pi,$ we use the Bloch theorem which makes it possible to restrict to a configuration space   
$x\in [0,2\pi[$ with periodic boundary conditions, changing only the kinetic energy term in the Hamiltonian $p^2/2$ to $(p+\hbar\beta)^2/2,$ where $\beta \in ]\!-\!1/2,1/2]$ is the Bloch vector. 	
The configuration space, $x\in [0,2\pi[$, is discretized in $N$ equidistant points; in momentum space, this corresponds to wavevectors (that is, up to multiplicative factor $\hbar$, momenta) in the $]\!-\!N/2,N/2]$ range. Switching from the configuration space representation of the wavefunction to the momentum space representation involves a Fourier transform of length $N,$ (the dimension of the Hilbert space) which can be done efficiently.

Altogether, the propagation algorithm is thus a series of forward and backward Fourier transforms interleaved with multiplication of each component
of the current state by a phase factor. The initial state is chosen as a $\delta$ function at the origin $\psi(p,t\!=\!0)=\delta(p).$ The quantity
$|\psi(p,t)|^2$ is thus the intensity propagator at time $t.$ The averaging over disorder realizations is performed firstly by averaging over many values of the Bloch vector $\beta,$ and secondly by averaging over the phases $\varphi_2,\varphi_3$ of the quasi-periodic kick amplitude modulation. After averaging $|\psi(p,t)|^2$ over the disorder realizations, we obtain the disorder-averaged intensity propagator $P_\mathrm{QPKR}(p,t)\equiv \overline{|\psi(p,t)|^2}.$ A simple rescaling of the momentum $p$ to $\mathcal{P}=pt^{-1/3}$ provides us with the quantity $\mathcal{N}(\mathcal{P},t)=t^{1/3}P_\mathrm{QPKR}(p,t).$ 

The size $N$ of the Hilbert space must be chosen sufficiently large for the momentum distribution to be negligibly small at the maximum momentum $|p|\!=\!N\hbar/2.$ We used up to $N=49152$ for the longest time considered $t\!=\!4\times 10^8.$ The averaging was performed over 17600 disorder realizations for times up to $t\!=\!10^6,$ 8800 for $t\!=\!10^7,$  1536 for $\!t=\!10^8$ and 120 for $t\!=\!4\times 10^8.$

\begin{figure}
\includegraphics[width=\columnwidth]{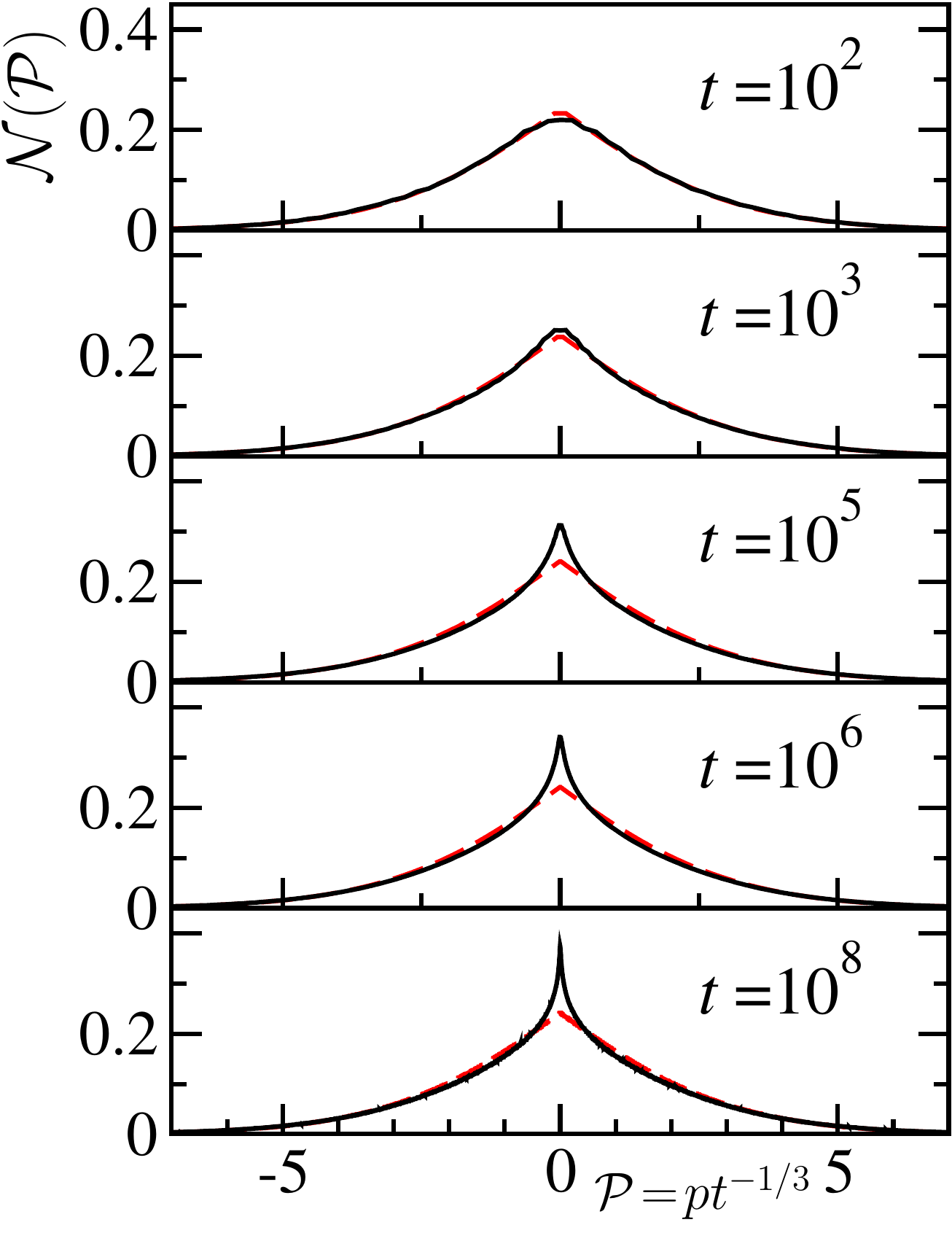}
\caption{Temporal evolution in momentum space of a wavepacket launched at $t\!=\!0$. The numerical simulation (black solid curve) is performed for the quasi-periodically kicked rotor at the critical point of the 3D Anderson transition. The global expansion is sub-diffusive in momentum space $\overline{\langle p^2(t)\rangle} \propto t^{2/3}.$ When plotted vs. dimensionless rescaled coordinate $\mathcal{P}=pt^{-1/3},$ the density in momentum space $\mathcal{N}$ takes a time-independent shape -- predicted to be a Airy function, Eq.~(\ref{Eq:fonccritAi}) (red dashed curve) -- except near the origin where a sharp peak grows with time. This peak is a direct manifestation of multifractality at the critical point. Parameters are $K=8.096,\varepsilon=0.4544,\omega_2=2.67220902067011,\omega_3=2.01719090924681,\hbar=3.54.$}
\label{Fig:naive}
\end{figure}

In Fig.~\ref{Fig:naive}, we show the numerically computed $\mathcal{N}(\mathcal{P})$ at various times (number of kicks), right at the critical point of the Anderson transition. While the agreement with Eq. (\ref{eq:airyfunc})
is excellent at short time (100 kicks, comparable to the duration of the experiment), a sharp peak near $p=0$ develops at increasingly long times. The existence and properties of this peak is the central subject of this paper. We show below that this peak -- not described by the SCTL -- is a manifestation of multifractality
at the critical point and is directly related to the multifractal dimension $d_2.$

\section{Multifractal model of the sharp peak}
\label{Sec:Theory}
\subsection{3D disordered system}
To understand the origin of this sharp peak, it is easier to leave the QPKR for a moment and turn back to a standard
disordered system such as the Anderson model, where 3D localization takes place in configuration space.
The average expansion of a wavepacket with time is described by the disorder-averaged intensity propagator $P(\vecr_1,\vecr_2,t)$, which gives the average probability to move from $\vecr_1$ at $t=0$ to $\vecr_2$ at time $t$.
Its Fourier transform w.r.t. time $t$, the propagator 
\begin{equation}
P(\vecr_1,\vecr_2,\omega) =  \frac{\overline{G^R(\vecr_1,\vecr_2,E\!+\!\omega/2)G^A(\vecr_2,\vecr_1,E\!-\!\omega/2)}}{2\pi \nu}
\label{Eq:Intensity_Propagator}
\end{equation}
is proportional to the disorder average of the product of retarded and advanced 
Green's functions at energy $E$, with $\nu$ the density of states per unit volume. Thanks to the statistical translational invariance of the disorder, it depends only on 
$\vecr=\vecr_1-\vecr_2.$ For simplicity of the notations, we write only $\vecr$ and omit the $E$-dependence of quantities.
It is convenient to consider the temporal and spatial Fourier transform $P(\vecq,\omega)=\int{\rmd t\ \rmd^d \vecr\ P(\vecr,t)\ \rme^{i\omega t-i\vecq.\vecr}}.$ Because the disordered potential is statistically isotropic, it depends only 
on the modulus $|\vecq|$ and can be written as:
\begin{equation}
 P(\vecq,\omega) = \frac{1}{-i\omega + D(q,\omega)q^2}.
\label{Eq:Pq}
\end{equation}
Eq. (\ref{Eq:Pq}) defines the momentum and frequency dependent diffusion coefficient $D(q,\omega)$. Causality implies that $P(\vecr,t)$ vanishes for negative $t$, while unitarity of the Hamiltonian evolution implies the conservation of probability $\int{\rmd^d \vecr\ P(\vecr,t)}=1$ for all $t>0,$ i.e.:
\begin{equation}
P(q\!=\!0,\omega) = \frac{1}{-i\omega}.
\end{equation} 
Causality also implies that $D(q,\omega)$ has no singularity in the upper complex half-plane $\Im \omega>0$ and cannot diverge more rapidly than $1/q^2$ at small $q.$
Because  $P(\vecr,t)$ is a real function, its Fourier transform satisfies:
\begin{equation}
P(q^*,\omega^*) = P^*(q,-\omega) 
\end{equation} 
where the $^*$ denotes complex conjugation, so that
\begin{equation}
D(q^*,\omega^*) = D^*(q,-\omega). 
\end{equation} 
In particular, $D(q,i\omega)$ must be real for real $q,\omega.$ 

Of particular interest is the small $\omega$ limit of $D(q,\omega)$, which describes long times.
In this limit and in a usual diffusive system, $D(q,\omega)\!=\!D_0\!=\!\ell^2/(3\tau)$  equals the classical diffusion coefficient, where $\ell$ is the mean free path and $\tau\!=\!m\ell/\hbar k$  the mean free time. In a localized system, $D(q,\omega)\!=\!-i\omega\xi^2,$ with $\xi$ the localization length. At the critical point finally, the SCTL predicts $D(q,\omega)\propto (-i\omega)^{1/3}$~\cite{Shapiro:SCTL:PRB82,MuellerDelande:Houches:2009}, which yields Eq.~(\ref{Eq:fonccritAi}).
In turn, deviations from the Airy shape, as visible in Fig.~\ref{Fig:naive}, imply that $D(q,\omega)$ must deviate from the simple $(-i\omega)^{1/3}$ dependence. Following earlier analyses of the intensity propagator~\cite{Wegner:Critical:ZPB76,Wegner:Critical:Book85}, Chalker~\cite{Chalker:Multifractality:PhysicaA1990} proposed that, at short distance (large $q$), 
$D(q,\omega)$ acquires a non-trivial $q$-dependence that we now recall.
$D(q,\omega)$ must respect  the one-parameter scaling law characterizing the Anderson transition at large distance and long time~\cite{Abrahams:Scaling:PRL79}. This scaling law involves the following relevant length scales:  the mean free path $\ell$,
$1/q,$ and  $L_{\omega}\!=\!\ell (\omega \tau)^{-1/3},$ the mean distance traveled by a particle in time $1/\omega$ at the critical point.
The localization length $\xi$ is in general an additional characteristic length, but at the critical point it is infinite and thus irrelevant.
In the following, we will only consider the long time limit $\omega \tau \ll 1,$ so that
$L_{\omega} \gg \ell.$
The one-parameter scaling law implies that $q$ appears in $D(q,\omega)$ only through
the $qL_{\omega}$ combination. Under this constraint, Chalker's ansatz \footnote{In ~\cite{Chalker:Multifractality:PhysicaA1990}, a fourth regime, labeled (iii), is introduced between the normal and the multifractal regimes, with $D(q,i\omega)\propto q$. We tried to add it in our fits, but the quality of the results was much worse. We thus conclude that this regime does not actually exist.} distinguishes the three following regimes~\cite{Chalker:Multifractality:PhysicaA1990,Brandes:Multfractality:ADP1996}:
\begin{itemize}
 \item[(A)] When $qL_{\omega}\!<\!1$ (long distance), multifractal correlations have no time to develop and one expects the normal sub-diffusive behavior, i.e. $D(q,i\omega) \propto D_0 (\omega\tau)^{1/3}.$ This is region (ii) in~\cite{Chalker:Multifractality:PhysicaA1990}.
 \item[(B)] When $qL_{\omega}\!>\!1$ (short distance), but still $q\ell\!<\!1,$ multifractality sets in and $D$ takes a $q$ dependence: $D(q,i\omega)\!\propto \!D_0 (\omega\tau)^{1/3} (qL_{\omega})^{d_2-2}.$ This is region (v) in~\cite{Chalker:Multifractality:PhysicaA1990}.
 \item[(C)] Finally, at very short distance $q\ell\!>\!1$, the mean free path sets a non-universal
 cutoff ensuring that the propagator $P(q,\omega)$ falls off sufficiently rapidly at large $q,$ so that no unphysical
 singularity exists in $P(\vecr,t)$ below the mean free path. While (A) and (B) obey the one-parameter scaling law, regime (C) breaks it at short distance where no (sub)-diffusive behavior makes sense. This is the extreme left part of region (v) in~\cite{Chalker:Multifractality:PhysicaA1990}.
\end{itemize}

\subsection{Singularity of the propagator near the origin in 3D disordered systems}
The behavior of the disorder-averaged intensity propagator (\ref{Eq:Intensity_Propagator}) near the origin $\vecr=0$ is a bit subtle at the critical point. 
We first consider the non-multifractal case -- regime (A) -- where the diffusion coefficient $D(q,\omega)$ scales like $\omega^{1/3}.$
We use the mixed momentum-time representation of the intensity propagator:
\begin{equation}
P(\vecq,t) = \int{\rmd^d \vecr\ P(\vecr,t)\ \rme^{-i\vecq.\vecr}} =  \int{\frac{\rmd \omega}{2\pi}\ P(\vecq,\omega)\ \rme^{-i\omega t}} 
\label{Eq:Mixed}
\end{equation}
At very large $q$, the $-i\omega$ term in the denominator of Eq.~(\ref{Eq:Pq}) can be neglected and the integral over $\omega$ computed exactly, e.g. using Eq. 3.761.9 in~\cite{Gradshteyn:1994}. The result is $\propto q^{-2}t^{-2/3}.$ The $1/q^2$ behavior at large $q$ converts, after a 3D Fourier transform, into a $1/r$ divergence in configuration space:
\begin{equation}
P(\vecr,t) \propto t^{-2/3}r^{-1}.
\label{Eq:P3D_nonmultifractal}
\end{equation}

When the multifractal regime (B) comes into play -- that is at short distance $r$ -- $D(q,\omega)$ scales like $q^{d_2-2}\omega^{1-d_2/3}$ at large $q.$ 
Again, the  $-i\omega$ term in the denominator of Eq.~(\ref{Eq:Pq}) can be neglected and the integral over $\omega$ computed exactly.
The result is $\propto q^{-d_2}t^{-d_2/3}.$ After a 3D Fourier transform, this gives:
\begin{equation}
P_{\mathrm{multifractal}}(\vecr,t) \propto t^{-d_2/3} r^{d_2-3}.
\label{Eq:Pmult3D}
\end{equation}
Finally, at very short distance $r\approx \ell$ -- regime (C) -- the divergence of Eq.~(\ref{Eq:Pmult3D}) is smoothed out.

Numerical simulations on the Anderson model~\cite{Huckestein:Multifractality:PRB1999} and related  models~\cite{GarciaGarcia:Multifractality:PRL05,Kravtsov:ChalkerAnsatz:PRB2010}
have confirmed such a behavior and the existence of the regimes (A-C) and the associated scalings, especially non-integer algebraic exponents.

\subsection{Singularity near the origin for the QPKR}
How can we relate the laws (A-C) to the observed behavior for the QPKR? 
As discussed in section~\ref{sec:QPKR}, the dynamics of the 1D QPKR can be mapped onto the one of a 3D disordered system, with a class of specific wavefunctions, Eq.~(\ref{Eq:Psi3}), perfectly well localized along directions $x_2$ and $x_3,$
that is with a uniform density along the conjugate variables $p_2$ and $p_3$ at all times, and density scaling like $|\psi(p_1,t)|^2$ along the ``physical'' momentum $p=p_1.$  Thus, the
1D propagator of the QPKR simply follows from the 3D intensity propagator by summing over the transverse directions $p_2$ and $p_3.$ In Fourier space, this means that only 
$q_2\!=\!q_3\!=\!0$ contributes:
\begin{eqnarray}
P_\mathrm{QPKR}(p,t)\!&=&\!\int{\frac{\rmd\omega}{2\pi}\rme^{-i\omega t}\!\int{\frac{\rmd q}{2\pi}\rme^{iqp}}P(q,q_2\!=\!0,q_3\!=\!0,\omega)}\nonumber\\
&=&\iint{\frac{\rmd\omega\ \rmd q}{4\pi^2}\ \frac{\rme^{i(qp-\omega t)}}{-i\omega+D(q,\omega)q^2}}.
\label{Eq:FT_QPKR}
\end{eqnarray}
Alternatively, the 1D intensity propagator could be obtained by integrating the 3D intensity propagator over the two transverse directions: $\iint{P(x,y,z,t)\ \rmd y\ \rmd z}.$

At this stage, let us point a slight complication: we so far assumed the 3D disordered system to be statistically isotropic. 
In~\cite{Lemarie:KR3DClassic:JMO10} however, it has been shown that the 3D Anderson system on which the QPKR is mapped is anisotropic so that the 
disorder-averaged propagator should be strictly speaking written as:

\begin{equation}
P(\vecq,\omega) = \frac{1}{-i\omega + \vecq.\vecD(\vecq,\omega).\vecq},
\end{equation}
where $\vecD$ is the anisotropic diffusion tensor. 
Because the 1D propagator of the QPKR involves only $q_2\!=\!q_3\!=\!0$ though, the only component of $\vecD$ that matters is $D_{11},$ so that everything boils down to Eq.~(\ref{Eq:FT_QPKR}).

The Fourier transform from $\omega$ to $t$ is eventually identical for the QPKR and for a disordered 3D system. Thus, the mixed representation $P(q,t),$ Eq.~(\ref{Eq:Mixed}), is identical in both cases. It is only the 3D or 1D Fourier transform
w.r.t. the $q$ variable which makes a difference.

We have now all the ingredients at hand to infer $\mathcal{N}(\mathcal{P}\!=\!pt^{-1/3},t)\!=\!t^{1/3}P_\mathrm{QPKR}(p,t)$ for the QPKR.  We first look at the non-multifractal contribution, assuming that the $1/q^2t^{2/3}$ behavior at large $q$ -- regime (A) -- is valid everywhere. A simple 1D Fourier transform converts it in
a $|p|/t^{2/3}$ singularity in the intensity propagator $P_\mathrm{QPKR}(p,t)$. 
At $p=0,$ there is a constant contribution scaling like $t^{-1/3},$ in accordance with the one-parameter scaling law, finally leading to 
\begin{equation}
\mathcal{N}(\mathcal{P}) \approx \alpha -\beta|\mathcal{P}|
\end{equation}
at small $\mathcal{P}=pt^{-1/3}.$ In fact, it is possible to perform exactly the full double Fourier transform, see~\cite{Lemarie:These:09,Lemarie:CriticalStateAndersonTransition:PRL10}. The result is Eq.~(\ref{eq:airyfunc}), which displays explicitly the expected linear singularity near the origin~\footnote{An alternative approach, using fractional diffusion equations, can be used to compute the intensity propagator both in the non-multifractal and multifractal regimes. It produces of course the same results. This will be the subject of a forthcoming publication.}.

When the multifractal regime comes into play, $D(q,\omega)$ scales like $q^{d_2-2}\omega^{1-d_2/3}$ at large $q$ -- regime (B) --
resulting in $P(q,t) \propto q^{-d_2}t^{-d_2/3}.$ The 1D Fourier transform gives a $|p|^{d_2-1}/t^{d_2/3}$ singularity near $p=0,$ that is  $P_\mathrm{QPKR}(p,t)\!=\! t^{-1/3} \left( \alpha - \beta |p t^{-1/3}|^{d_2-1}\right)$ or: 
\begin{equation}
\mathcal{N}(\mathcal{P})= \alpha - \beta |\mathcal{P}|^{d_2-1},
\label{Eq:Pmult1DKR}
\end{equation}
at small $\mathcal{P}$.

The $\alpha,\beta$ constants are not universal and depend on the boundary around $qL_{\omega}\!=\!1$ between the normal sub-diffusive and the multifractal regions (A) and (B), but the algebraic dependence $|p t^{-1/3}|^{d_2-1}$ is universal. Note that because $d_2 \approx 1.24$ at the critical point of the 3D Anderson transition~\cite{Rodriguez:MultifractalFiniteSizeScaling:PRB11}, the 3D intensity propagator, Eq.~(\ref{Eq:Pmult3D}), has an algebraic divergence near $r\!=\!0$ while the 1D intensity propagator of the QPKR is finite at $p\!=\!0$, with a non-integer power law singularity.

Whereas the contribution of the mean-field regime (A) to $D(q,\omega)$ leads to the kink $\mathcal{N}(\mathcal{P})\!-\!\mathcal{N}(0) \propto |\mathcal{P}|$ at small $\mathcal{P},$ Eq.~(\ref{eq:airyfunc}), the multifractal law (\ref{Eq:Pmult1DKR}) is more singular: it is responsible for the
small peak near the origin observed in Fig.~\ref{Fig:naive}.
At short time (say shorter than 100 kicks), this rather weak singularity at the origin
is cut at the mean free path, and the normal component (the Airy function) reproduces
very well the numerical calculation. As time grows, the whole wavepacket
spreads in size like $t^{1/3}$ making the short distance cutoff to act at
smaller and smaller $\mathcal{P}=pt^{-1/3}.$ Because $d_2>1,$ the algebraic term in Eq.~(\ref{Eq:Pmult1DKR}) does not diverge at $\mathcal{P}\!=\!0,$ only its derivative is infinite.

\section{Extraction of the multifractal dimension $d_2$}
\label{Sec:Measurement} 
We can now use the numerically computed wavepackets to extract the value of the multifractal dimension $d_2.$ We have used two different methods, see Fig.~\ref{Fig:fit_comparison}. 

\subsection{First fitting procedure}
\begin{figure}
	\includegraphics[width=0.9\columnwidth]{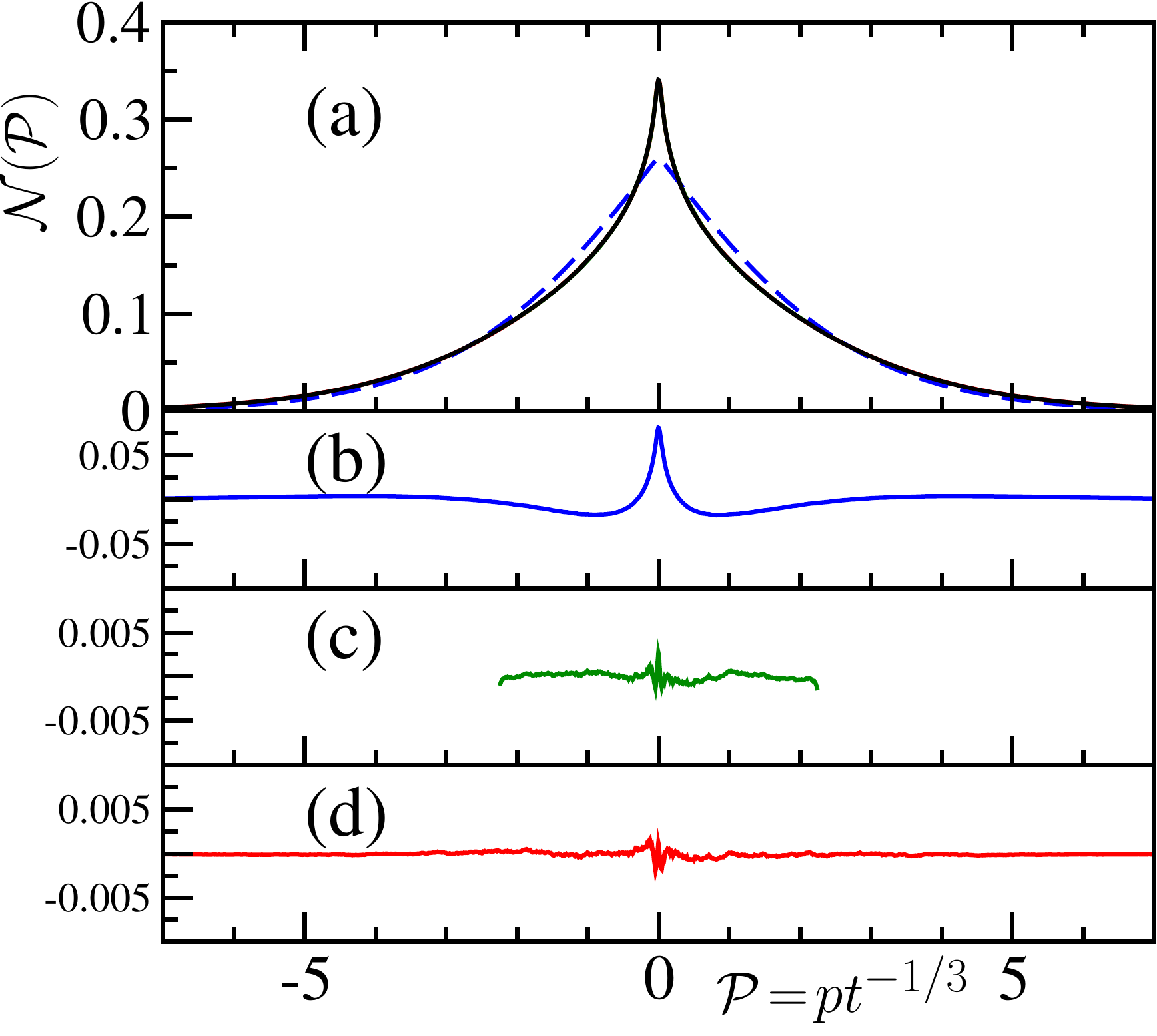}
	\caption{(a) Solid black curve: Numerically computed temporal evolution at $10^6$ kicks) in momentum space of a wavepacket launched at $t\!=\!0$ near momentum $p\!=\!0$ (see Fig.~\ref{Fig:naive} for the parameter values). The fit by an Airy 
		function, Eq.~(\ref{Eq:fonccritAi}), dashed blue curve, which does not take into account multifractality, is obviously bad; The residual (difference between the curve and the best fit) is shown in (b). The central region is very well fitted by an algebraic dependence, Eq.~(\ref{Eq:Pmult1DKR}),
		(residual shown in (c)) and gives $d_2=1.28\pm0.03$ (note the vertical scale 10 times smaller than in (b)). A fit of the full numerical curve interpolating between the three regimes (A-C) and in particular incorporating the multifractal regime (B) is indistinguishable from the numerical result, and gives $d_2=1.28\pm0.02$; The residual is shown in (d). 
		\label{Fig:fit_comparison}}
\end{figure}

The first fitting procedure uses only the very central part, near $\mathcal{P}=0,$ of the numerically computed disorder-averaged intensity propagator. 
Indeed, Eq.~(\ref{Eq:Pmult1DKR}) predicts an algebraic cusp at small $\mathcal{P}$ 
clearly visible at long times.
We thus fitted the central part of the numerically computed $\mathcal{N}(\mathcal{P},t)$ with Eq.~(\ref{Eq:Pmult1DKR}), with three fitting parameters
$\alpha,\beta$ and $d_2.$  The range of $\mathcal{P}$ values used must be not too large, as the fitting expression is expected to be valid only near $\mathcal{P}=0.$ We chose to include points up to $|\mathcal{P}|=2.25$ -- see Fig.~\ref{Fig:fit_comparison}(c) -- but the extracted $d_2$ value turns out to depend only weakly on the range used. This simple procedure already gives very satisfactory results, with values of $d_2$
almost independent of time at long time, although a separate fit is done for each time. 

At very short $p,$ of the order of the mean free path, the disorder-averaged intensity propagator does not obey the one-parameter scaling law of the Anderson transition (regime (C)), so that the expression (\ref{Eq:Pmult1DKR}) is not expected to be valid. In other words, the algebraic cusp at small $p$ is smoothed over one mean free path. The corresponding range in $\mathcal{P}=pt^{-1/3}$ shrinks when $t$ increases, explaining why the peak near $p=0$ grows. Such a smoothing affects the quality of the fit. In order to take this fact into account, we have to exclude a small region around $\mathcal{P}=0$ from the fit. The results are essentially independent of the size of this small region.

\subsection{Second fitting procedure}
The second fitting procedure uses the full numerically computed disorder-averaged intensity propagator. It assumes that the momentum-frequency dependent diffusion coefficient follows the Chalker's ansatz~\cite{Chalker:Multifractality:PhysicaA1990} with the three different regimes presented in the main text. More precisely, we use the following ansatz:
\begin{equation}
D(q,i\omega) = \frac{3}{2^{2/3}}\ D_0\ (\omega \tau)^{1/3}\ f(qL_\omega) 
\label{Eq:DChalker}
\end{equation}
where $\tau$ is the mean scattering time, $D_0=\ell^2/3\tau$ the classical Boltzmann diffusion coefficient ($\ell$ is the mean free path) and
$L_\omega=\ell(\omega\tau)^{1/3}$ is the mean distance traveled at the critical point in time $1/\omega.$
The fact that the real function $f$ depends only on the product $qL_\omega$ is a requirement of the one-parameter scaling law. In the non-multifractal regime (A) where $qL_\omega\ll 1,$ the self-consistent theory of localization predicts that $f$ is constant~\cite{Lopez:PhaseDiagramAndersonQKR:NJP13}.
The precise constant value of $f$ depends on the cutoffs used in the self-consistent theory~\cite{Vollhardt:SelfConsistentTheoryAnderson:92,Lopez:PhaseDiagramAndersonQKR:NJP13,MuellerDelande:Houches:2009}. If the cutoff is chosen so that the transition takes place at $k\ell=1,$ the numerical factors in Eq.~(\ref{Eq:DChalker}) are such that $f=1$ in the non-multifractal regime.

In the multifractal regime $qL_\omega>1,$  the Chalker's ansatz~\cite{Chalker:Multifractality:PhysicaA1990} states that $D(q,i\omega)$ 
scales like $q^{d_2-2}\omega^{1-d_2/3}$ or, equivalently, $f(qL_\omega)\propto (qL_\omega)^{d_2-2}.$ There are of course many possibilities to smoothly connect the $f(x)=1$ behavior at small $x$ to the $f(x)\propto x^{d_2-2}$ decrease at large $x$. The only requirement is that the transition between the two regimes takes place around $x=1.$ In order to avoid unphysical Gibbs-like oscillations after Fourier transform, we used the following smooth ansatz:
\begin{equation}
f(x) = \left[1+(x/x_0)^{\gamma (2-d_2)}\right]^{-1/\gamma}
\end{equation} 
where $\gamma$ is a positive exponent and $x_0$ a number of the order of unity characterizing the transition point between the two regimes.
This ansatz is of course a bit arbitrary. We have tried a few other ways of smoothly connecting the two regimes, which give very similar final results.
When the parameters $D_0,d_2,x_0,\gamma$ and $\tau$ (or equivalently $\ell$) are given, $D(q,i\omega)$ is entirely specified. In order to compute the disorder-averaged intensity propagator, one has to compute $D(q,\omega)$ for real $\omega$, which is rather easy by analytic continuation in the complex plane, as there is no singularity in the upper half-plane $\Im \omega>0.$ The last step is a double Fourier transform from $q,\omega$ to $p,t$ to obtain $P_\mathrm{QPKR}(p,t).$ In order to take into account the non-universal behavior at very short distance -- regime (C) -- we convoluted the obtained $P_\mathrm{QPKR}(p,t)$ by a Gaussian:
\begin{equation}
g(p)= \frac{1}{\sigma \sqrt{2\pi}}\ \exp\left(-\frac{p^2}{2\sigma^2}\right)
\label{Eq:Gaussian}
\end{equation}  
where $\sigma$ is a constant of the order of the mean free path $\ell.$  

We used these distributions to fit the numerical results obtained
for the kicked rotor. There are 5 different fit parameters: the first one is the ``classical'' diffusion coefficient $D_0$ which determines the overall scaling factor of the distribution (or equivalently the value of $\overline{\langle p^2(t)\rangle}/t^{2/3}$ at long time). Taking $D_0$ as a fit parameter accounts for the somewhat arbitrary numerical prefactor in Eq. (\ref{Eq:DChalker}), not accurately predicted by the self-consistent theory. 
The second fit parameter is the short distance cutoff $\sigma$ in Eq. (\ref{Eq:Gaussian}), of the order of the mean free path.  It turns out that the final results are essentially insensitive to the exact value of this parameter. 
The three left important parameters are $d_2$ (the figure of merit of our analysis), and $x_0,\gamma$ which describe the transition between
the normal and multifractal regimes for $D(q,\omega).$ We performed three fitting runs:
\begin{itemize}
	\item In the first run, we fitted all five parameters $D_0,\tau,d_2,x_0,\gamma$ for each time. We observed that the values of $x_0$ and $\gamma$ fluctuate in not too large intervals, that is $x_0 \in [0.24,0.40]$ and $\gamma \in [2.8, 4.0]$.
	\item In a second run, we fixed $\gamma$ at its most probable value $\gamma=3.0$ and fitted the remaining four parameters.
	\item In a third run, we additionally fixed $x_0$ at its most probable value $x_0=0.3$ and fitted the remaining three parameters.
\end{itemize} 
The results of the three fitting runs are very similar. Importantly, the residuals of the fits -- deviations between the numerical data and the fitting functions -- are very comparable for the three runs, so that they are of almost equal significance. The fluctuations of the fitted values for the three runs give an estimate of the error due the imperfections of the fits. Combined with the statistical uncertainty of the fit, they provide a reasonable estimate of the error bars on the determined values of $d_2.$

\subsection{Results}
\begin{table}[h!]
	\begin{center}
		\begin{tabular}{|c|c|c|c|c|c|c|c|c|}
			\hline	
			Time $t$& $10^3$  & $10^4$ & $10^5$ & $10^6$ & $10^7$ & $10^8$ & $4\!\times\!10^8$ \\
			\hline
			$d_2$ & $1.37$  & $1.33$ & $1.31 $ & $1.28$ & $1.26$ & $1.26$ & $1.26$  \\ 
			\hline
			$\Delta d_2$& $0.15$  & $0.05$ & $0.04$ & $0.03$ & $0.02$ & $0.02$ & $0.02$  \\
			\hline

		\end{tabular}
	\end{center}
	
	\caption{\label{Table_algebraic} Multifractal exponent $d_2,$ with estimated uncertainty $\Delta d_2,$ extracted using the first fitting procedure, that is from fits of the disorder-averaged intensity propagator near momentum $p=0$ to Eq.~(\ref{Eq:Pmult1DKR}), for various times $t$. The uncertainty is not the statistical error of the fit, but rather reflects the fluctuations of the result of the fit when the momentum range and the short-range cutoff are varied. Nevertheless, the result at long times is remarkably stable, proving the robustness of the fitting procedure.}
\end{table}

\begin{table}[h!]
	\begin{center}
		\begin{tabular}{|c|c|c|c|c|c|c|c|c|c|}
			\hline	
			Time $t$& $10^2$  & $10^3$  & $10^4$ & $10^5$ & $10^6$ & $10^7$ & $10^8$ & $4\!\times\!10^8$ \\
			\hline
			$d_2$ & $1.19$  & $1.32$ & $1.34 $ & $1.33$ & $1.28$ & $1.25$ & $1.24$ & $1.24$ \\ 
			\hline
			$\Delta d_2$& $0.2$  & $0.2$ & $0.08$ & $0.04$ & $0.02$ & $0.015$ & $0.015$ & $0.01$ \\
			\hline

		\end{tabular}
	\end{center}
	
	\caption{\label{Table_full} Multifractal exponent $d_2,$ with estimated uncertainty $\Delta d_2,$ extracted using the second fitting procedure, that is from fits of the full disorder-averaged intensity propagator, for various times $t$. The uncertainty is the combination of the statistical error of the fit and of the three different values that are obtained when the additional parameters $x_0,\gamma$ are either fitted or fixed. In any case, the smallness of $\Delta d_2$
		as well as the quality of the fit, see Fig.~\ref{Fig:fit_comparison}, validates the Chalker's ansatz and proves that it is experimentally possible
		to measure the multifractal exponent $d_2.$ }
\end{table}
In Table~\ref{Table_algebraic}, we give the values of $d_2 $ extracted from the numerical data using the first fitting method, for various times. The uncertainties take into account the fluctuations of the results when the range of $\mathcal{P}$ used for the fit is varied. It consistently gives a value of $d_2$ in the $[1.24,1.37]$ range for a considerably large
time interval, between $10^3$ and $4\times 10^8$ kicks, in good agreement with the known value
$1.24\pm 0.015$ for the 3D Anderson model~\cite{Rodriguez:MultifractalFiniteSizeScaling:PRB11}. 
(a more accurate value 
$d_2=1.243\pm0.006$ is given in the unpublished thesis~\cite{Vasquez:Thesis:2010}).

With the second fitting method, we found that the fitted $d_2$
is almost insensitive to the details of the interpolation between the three regimes. The obtained values of $d_2$ are given in Table~\ref{Table_full}. They are more or less time-independent
at long time, which strongly supports the validity of the Chalker's ansatz. They also agree well with the results of Table~\ref{Table_algebraic} and with the known value~\cite{Rodriguez:MultifractalFiniteSizeScaling:PRB11}.

For $t\!=\!10^6$ kicks, the two fitting methods give almost identical results, $d_2=1.28,$ and the quality of the fits is excellent, as shown in Fig.~\ref{Fig:fit_comparison}. The Airy
function, in contrast, strongly deviates from the numerical result.

\begin{figure}
\includegraphics[width=\columnwidth]{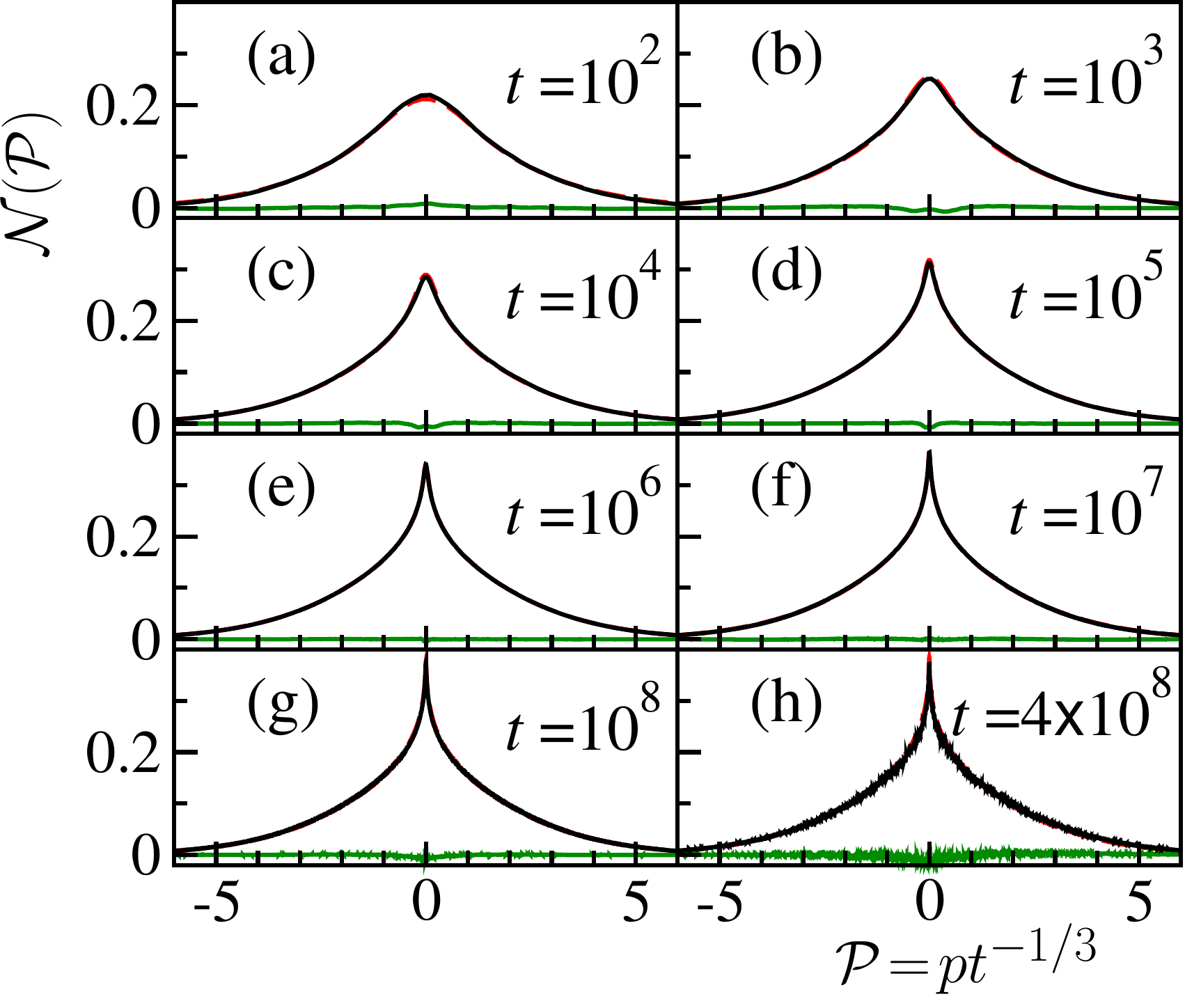}
\caption{Black solid lines: numerically computed temporal evolution in momentum space of a wavepacket launched at $t=0$ near momentum $p=0$ (see Fig.~\ref{Fig:naive} for the parameter values). Red dashed lines, often hidden by the black ones: prediction taking into account the sub-diffusive dynamics and the multifractality of the eigenstates (regimes (A-C)). The agreement is excellent (residuals are displayed as the lower green curves) over more than 6 decades of time.
The same value $d_2=1.26$ has been used for all plots.	\label{Fig:multiple_fit}}
\end{figure}

We finally show in Fig.~\ref{Fig:multiple_fit} that the  same value of $d_2$ allows us to reproduce almost perfectly the full momentum
distribution over a very wide range of times. The fact that a unique form of $D(q,\omega)$ 
reproduces the numerical results over more than 6 orders of magnitude of $t$ is on the one hand a very strong hint that the 
one-parameter scaling law remains valid for the Anderson transition in the multifractal regime, and on the other hand a confirmation of the validity of the Chalker's ansatz.

\section{Experimental perspectives and summary}
\label{Sec:Summary}
We have unveiled the existence of a sharp multifractality peak at the critical point of the Anderson transition. Based on the equivalence between the time-dependent QPKR and the 3D anisotropic Anderson model, we have also shown that the multifractal dimension $d_2$ of a critical 3D system can be extracted from the peak in the frame of a 1D experimental setup. Although this in principle requires to reach extremely long times, we stress that, even after $t=10^3$ kicks -- a value already reached in state-of-the-art experiments~\cite{Manai:Anderson2DKR:PRL15}-- a significant deviation
from the Airy shape is already visible. This opens the way to an experimental measure of multifractality properties using the atomic kicked rotor. 
The method is in no way restricted to the kicked rotor and could be used in other disordered systems~\cite{Faez:Multifractality:PRL09}. In a full 3D system, the average intensity propagator, Eq.~(\ref{Eq:Pmult3D}), is also sensitive to $d_2.$ If not all three dimensions of space are experimentally accessible, averaging over one or two dimensions still preserves the information on $d_2$, although the singularity is somewhat smoothed out.

Note that an apparently similar phenomenon, an enhanced return probability, has been recently experimentally observed on the kicked rotor~\cite{Hainaut:EnhancedReturnOrigin:PRL17}. It originates from the constructive interference between pairs of time-reversed paths for time-reversal invariant systems and is \emph{completely different} from the "multifractal" peak: it manifests itself on a  much shorter spatial scale of the order of the mean free path, that is in regime (C) where the one-parameter scaling law is violated. Moreover, it exists only for \emph{periodic} driving, as discussed in~\cite{Hainaut:EnhancedReturnOrigin:PRL17,Hainaut:CFS:NCM18} and is thus an unrelated phenomenon.

\acknowledgments

We thank G. Lemari\'e, J.C. Garreau, P. Szriftgiser, A. Mirlin and V.E. Kravstov for useful discussions.  This work was granted access to the HPC resources of TGCC under the allocation 2013-056089 made by GENCI (Grand Equipement National de Calcul Intensif) and to the HPC resources of The Institute for scientific Computing and Simulation financed by Region Ile de France and the project Equip@Meso (reference ANR-10-EQPX- 29-01).

\bibliographystyle{apsrev4-1}
\bibliography{artdatabase_v20}

\end{document}